\documentclass[a4paper,final]{appolb}
\usepackage{epsfig}

\newcommand{\sqs} {\mbox{\ensuremath{\sqrt{s}}}}
\newcommand{\MeV} {\mbox{\ensuremath{\mathrm{MeV / c^2}}}}
\newcommand{\The}{\mbox{\ensuremath{\Theta^+}}}
\newcommand{\Ximm}{\mbox{\ensuremath{\Xi^{--}(1862)}}}
\newcommand{\Xin}{\mbox{\ensuremath{\Xi(1862)}}}
\newcommand{\Thed}{\mbox{\ensuremath{\Theta^+ \rightarrow p K^0_s}}}
\newcommand{\pKns}{\mbox{\ensuremath{p K^0_s}}}
\newcommand{\nKp}{\mbox{\ensuremath{n K^+}}}
\newcommand{\Ximmd}{\mbox{\ensuremath{\Xi^{--}(1862) \rightarrow \Xi^- \pi^-}}}
\newcommand{\Xinn}{\mbox{\ensuremath{\Xi(1862) \rightarrow \Xi \pi}}}
\newcommand{\Xibb}{\mbox{\ensuremath{\Xi \pi}}}
\newcommand{\Thecd}{\mbox{\ensuremath{\Theta^0_c \rightarrow D^* p}}}
\newcommand{\Kspp}{\mbox{\ensuremath{K^0_s \rightarrow \pi^+ \pi^-}}}
\newcommand{\Lampp}{\mbox{\ensuremath{\Lambda \rightarrow p \pi^-}}}
\newcommand{\ALampp}{\mbox{\ensuremath{\overline{\Lambda} \rightarrow \overline{p} \pi^+}}}
\newcommand{\XiLpi}{\mbox{\ensuremath{\Xi^- \rightarrow \Lambda \pi^-}}}
\newcommand{\XiALpi}{\mbox{\ensuremath{\overline{\Xi}^+ \rightarrow \overline{\Lambda} \pi^+}}}
\newcommand{\Lamp}{\mbox{\ensuremath{\Lambda(1520) \rightarrow p K^-}}}
\newcommand{\ALamp}{\mbox{\ensuremath{\overline{\Lambda}(1520) \rightarrow \overline{p} K^+}}}
\newcommand{\Bdsig}{\mbox{\ensuremath{\mathcal{B} \cdot d\sigma / dy}}}

\begin{document}

\title{Search for Pentaquarks with HERA-B %
\thanks{Presented at Cracow Epiphany Conference on Hadron Spectroscopy 2005}%
}
\author{Joachim Spengler, for the HERA-B collaboration
\address{Max-Planck-Institut f\"ur Kernphysik, Postach 103989, 69029 Heidelberg, Germany}
}
\maketitle
\begin{abstract}
A large data set of proton-nucleus collisions at \sqs\ =41.6 GeV has been
searched for \The\  and \Xin\   pentaquark candidates. In $2 \cdot 10^8$ inelastic
events we find no evidence for narrow signals ($\sigma \approx$5~\MeV )
in the \Thed\  and \Xinn\  channels. Upper limits on production cross sections
at mid-rapidity and on ratios of production cross sections to those of well
established resonances are presented. 
\end{abstract}
  
\section{Introduction}

The possible discovery of an exotic baryon with at least five constituent quarks
in $\gamma$n reactions at low energies~\cite{LEPS} has initiated a very
active search for these so-called pentaquark states. Up to now, twelve
experiments~[1 - 12] have reported evidence for a narrow resonance \The\
with a mass near to 1540~\MeV\  decaying into \pKns\  or \nKp\ 
final states. In contrast to this, the higher mass states \Ximmd ~\cite{NA49}
and \Thecd ~\cite{H1} were each observed by only one experiment.

However, despite all these experimental data, this subject remains
controversial because an increasing amount of experiments report
negative results on searches for these states~[15 - 27]. 
Furthermore, the masses of the \The\ candidates show a large and
systematic spread. For recent reviews of the experimental
situation see~\cite{DZIERBA, KABA, HICKS} and references therein.

\section{Data sample}

HERA-B is a fixed-target experiment which studies collisions of
protons with the nuclei of atoms in target wires positioned in the
halo of HERA's 920 GeV proton beam. 
\begin{figure}[thb]
\epsfig{file=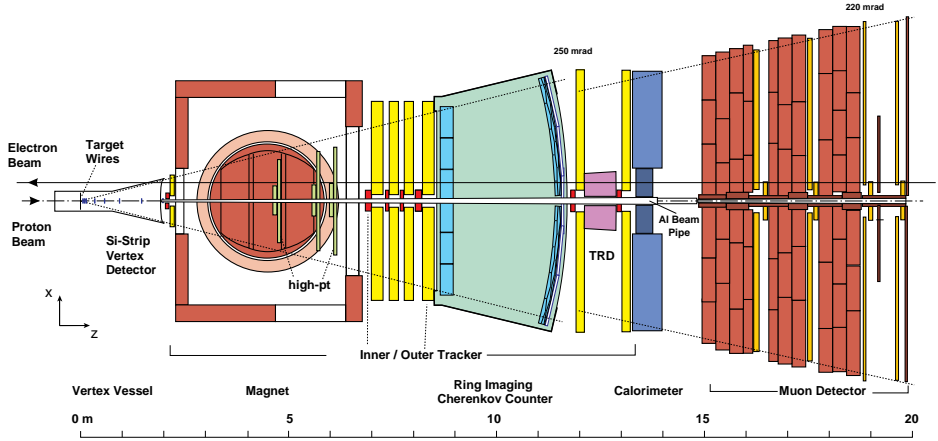,width=0.98\columnwidth}
 \caption{Plan view of the HERA-B detector.
    \label{fig:layout} }
\end{figure}
The large acceptance of the HERA-B 
spectrometer coupled with high-granularity particle-identification
devices and a precision vertex detector allow for detailed studies of
complex multi-particle final states. 
Fig.~\ref{fig:layout} shows a plan view of the detector in the
configuration of the 2002-2003 data run.

For this analysis~\cite{herab1} the information from the silicon vertex detector,
the main tracking system, the ring-imaging Cherenkov counter and the
electromagnetic calorimeter was used. A data
sample of $2 \cdot 10^8$ minimum bias events was recorded at
\sqs\ =41.6~GeV using target wires of different materials (carbon,
titanium and tungsten).

\begin{figure}[htb]
\epsfig{file=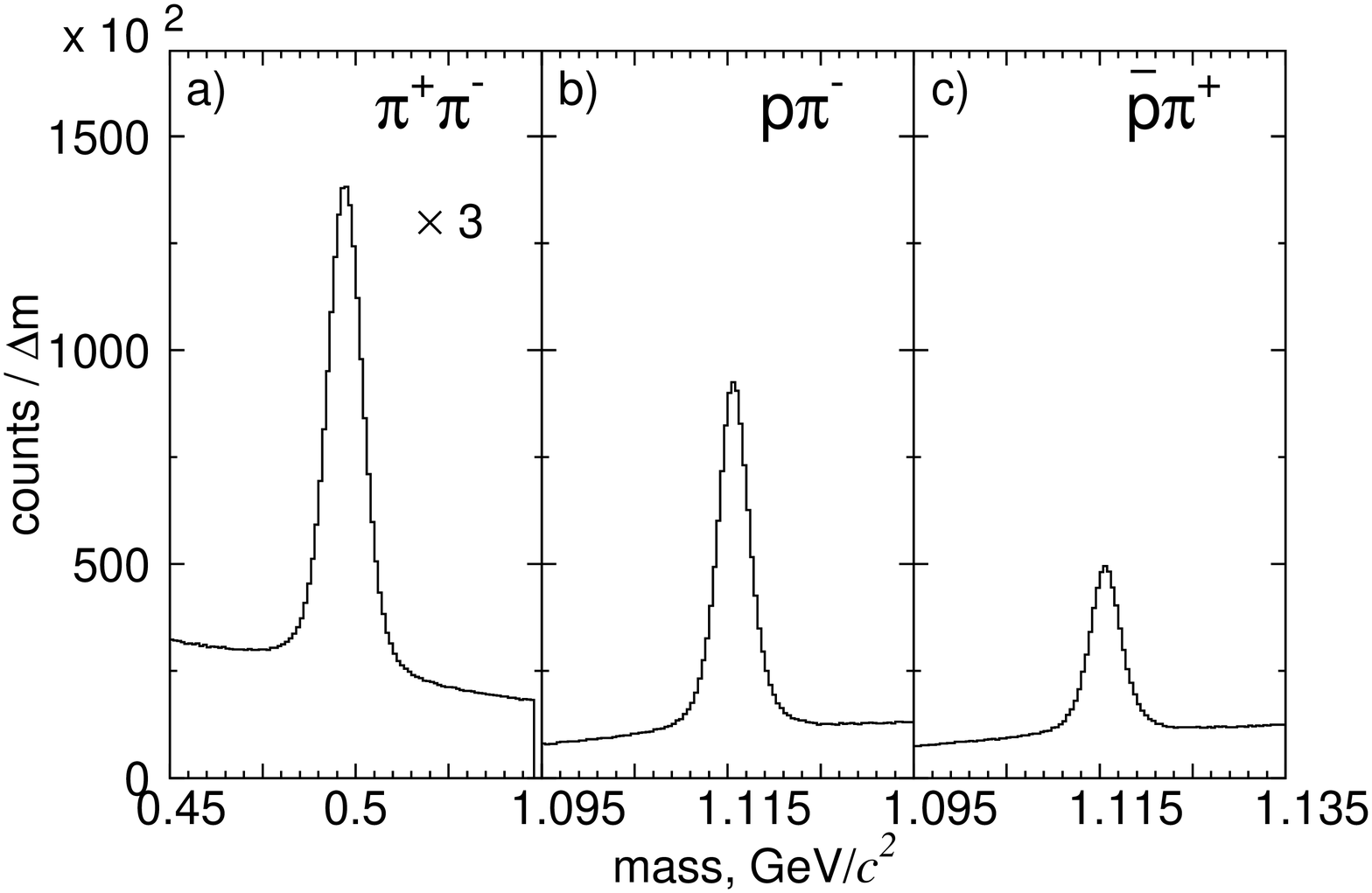,width=0.98\columnwidth}
\caption{Signals obtained with the C target from decays of
a) \Kspp , b) \Lampp\  and c) \ALampp .  
 \label{fig:V0plot} }
\end{figure}

\begin{figure}[ht]
\epsfig{file=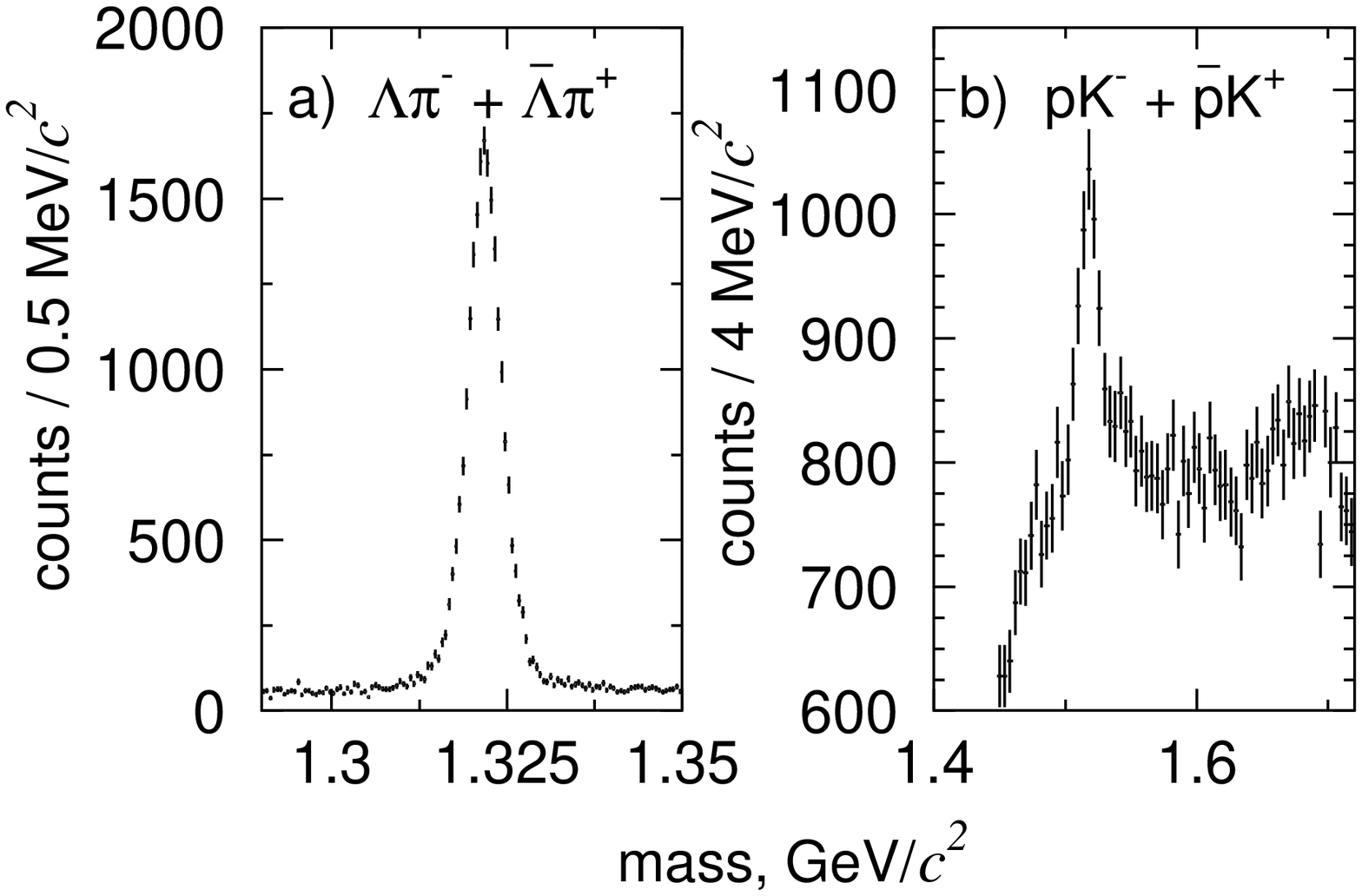,width=0.95\columnwidth}
\caption{Signals obtained with the C target from decays of a)
\XiLpi\  and \XiALpi , and b) \Lamp\  and \ALamp .  
    \label{fig:XiLa} }
\end{figure}

Applying soft vertex cuts, signals from \Kspp , \Lampp\  and
\ALampp\  are identified above a small background, as can be seen in
Fig.~\ref{fig:V0plot}.
$K^0_s / \Lambda$ ambiguities are rejected. Selecting $\Lambda$'s
by an invariant mass cut of 3$\sigma$ around the peak value and
requesting the $\Lambda \pi^-$ vertex
to be at least 2.5~cm downstream of the target, we obtain clean
signals of \XiLpi\  and charge conjugated (c.c.) decays
(see Fig.~\ref{fig:XiLa}a). 
The statistics of these signals together with their mass resolutions
is given in Table~\ref{tab:Resonan}.


\begin{table}[ht]
\caption{Statistics and experimental resolutions $\sigma$ of
the relevant signals. \label{tab:Resonan}}
\begin{center}
\begin{tabular}{|l|c|c|c|}
\hline
Signal &  C target & all targets & $\sigma$ / (\MeV )
\\ \hline \hline
$K^0_s$     & 2.2M   & 4.9M   &   4.9
\\ \hline
$\Lambda$ [c.c] & 440k [210k]   & 1.1M [520k]   &   1.6
\\ \hline
$\Lambda(1520)$ [c.c] & 1.3k [760] & 3.5k [2.1k]&   2.3
\\ \hline
$\Xi^-$ [c.c] & 4.7k [3.4k] & 12k [8.2k]        &   2.6
\\ \hline
$\Xi(1530)^0$ [c.c] & 610 [380] & 1.4k [940]    &   2.9
\\ \hline
\end{tabular}
\end{center}
\end{table}

\section{Search for \Thed\ }

Protons are identified by requesting that the proton likelihood, which is provided
by the RICH, be above 0.95. This cut allows to reduce the misidentification
probability to less than 1$\%$ in the selected momentum range from 22 to
55 GeV/c. As a reference state we reconstruct the $\Lambda(1520)$ by its
decay into $p K^-$. A prominent $\Lambda(1520)$ + c.c signal can be seen in
Fig.~\ref{fig:XiLa}b.

\begin{figure}[htb]
\epsfig{file=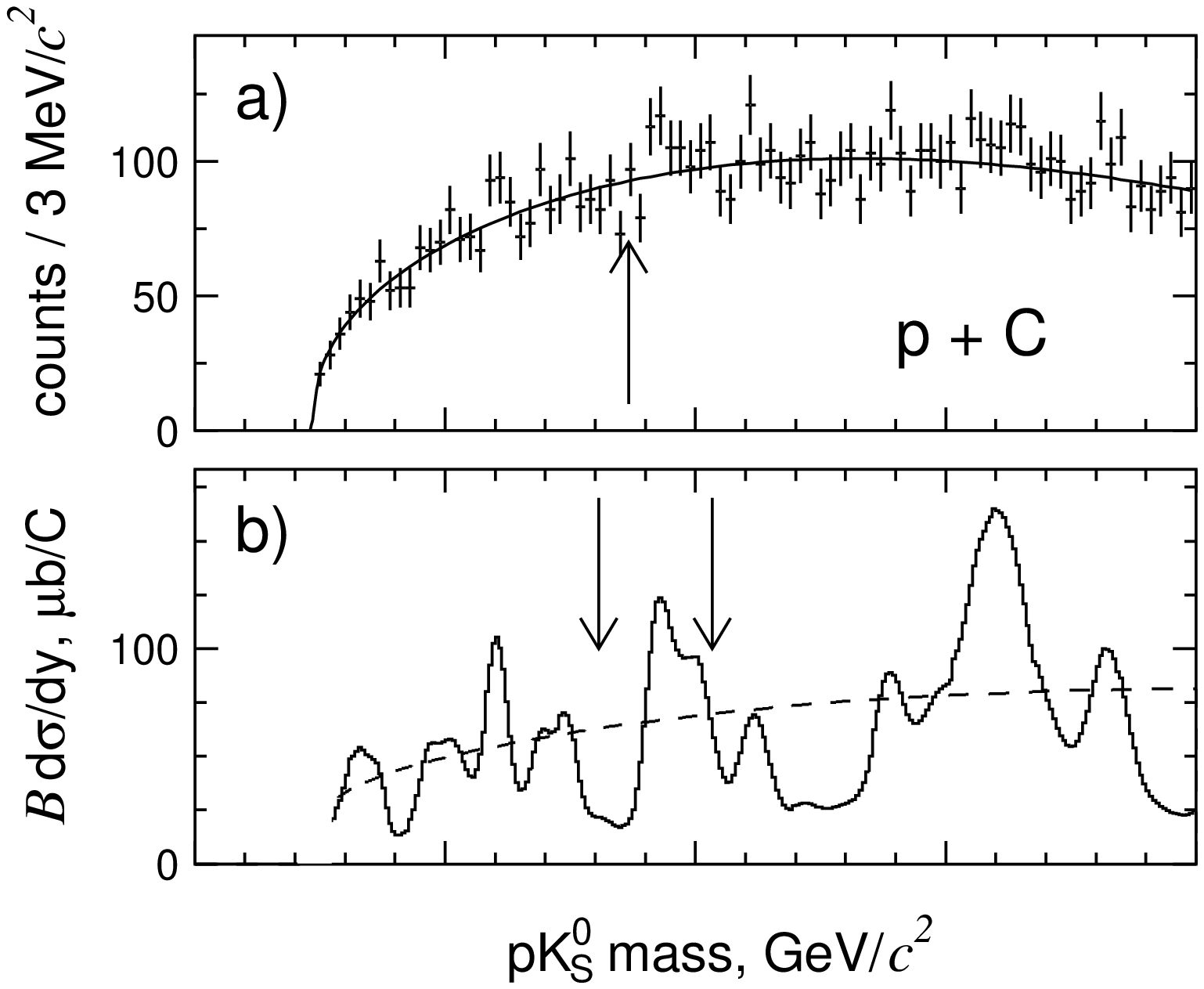,width=0.49\columnwidth}
\epsfig{file=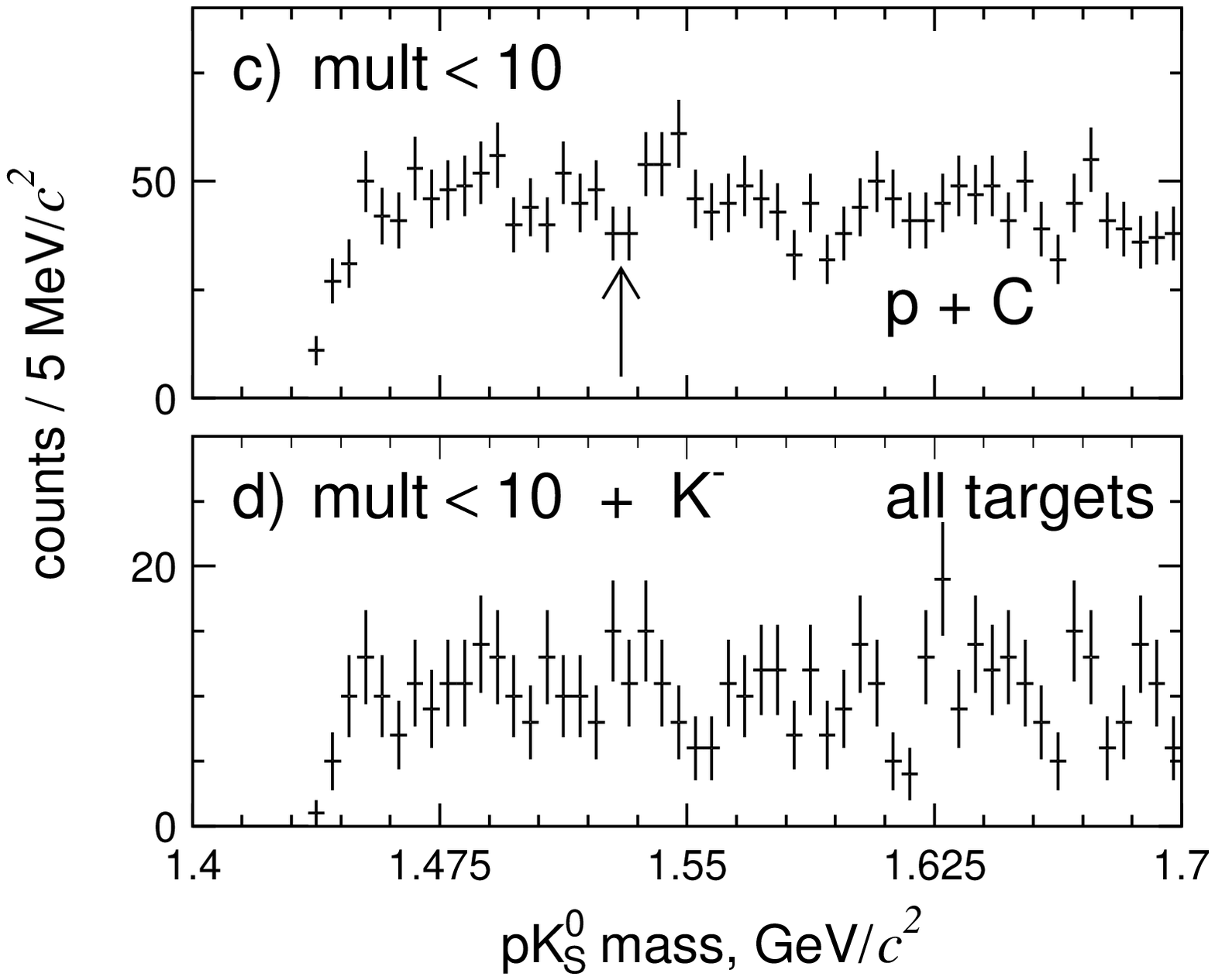,width=0.49\columnwidth}
\caption{The  $p K^0_s$ invariant mass distributions: a) data from pC collisions
and background estimate (line); b) upper limit at 95$\%$ CL
on \Bdsig\ per carbon nucleus; the dashed line shows our 95$\%$ CL sensitivity;
c), d) same as a) but requiring c) a charged track multiplicity of $<$ 10, and d)
in addition a strange particle in the event. The arrows mark the masses of
1521, 1530 and 1555~\MeV  .
\label{fig:prK0s} }
\end{figure}

For the search for \Thed\  decays, we combine these well identified
protons with $K^0_s$ candidates.
The invariant mass distribution is shown in Fig.~\ref{fig:prK0s}a for
the carbon sample. The spectrum exhibits a smooth shape which is well described
by the background estimate obtained from event mixing. In order to determine
an upper limit on the production cross section at mid-rapidity, we fit the
spectrum with a Gaussian plus a background of fixed shape. Due to the
uncertainty in the \The\  mass, the mean of the Gaussian is varied in steps of
1~\MeV . The resolution increases from 2.6 to 6.1~\MeV\  over the
considered mass range. The resolution is 3.9~\MeV\  at 1530~\MeV\  which
is approximately the mean of the reported \The\  masses in the \pKns\ 
channel. Using the prescription of Feldman and Cousins ~\cite{Feld}, we
arrive at the upper limit (95$\%$) curve for \Bdsig\ per carbon nucleus shown
as solid line in Fig.~\ref{fig:prK0s}b. The rapidity interval selected is
$y_{cm}=\pm 0.3$. The dashed line indicates the experimental sensitivity.
The range of reported mass values of \The\  candidates is indicated by arrows
in Fig.~\ref{fig:prK0s}b.

Further search strategies are tried including i) a cut on the charged track
multiplicity of $<$ 10 (Fig.~\ref{fig:prK0s}c), or ii) the additional request of a
strange tagging particle such as $\Lambda, \Sigma$ or $K^-$ (Fig.~\ref{fig:prK0s}d).
No statistically significant structure can be observed.

\begin{figure}[htb]
\epsfig{file=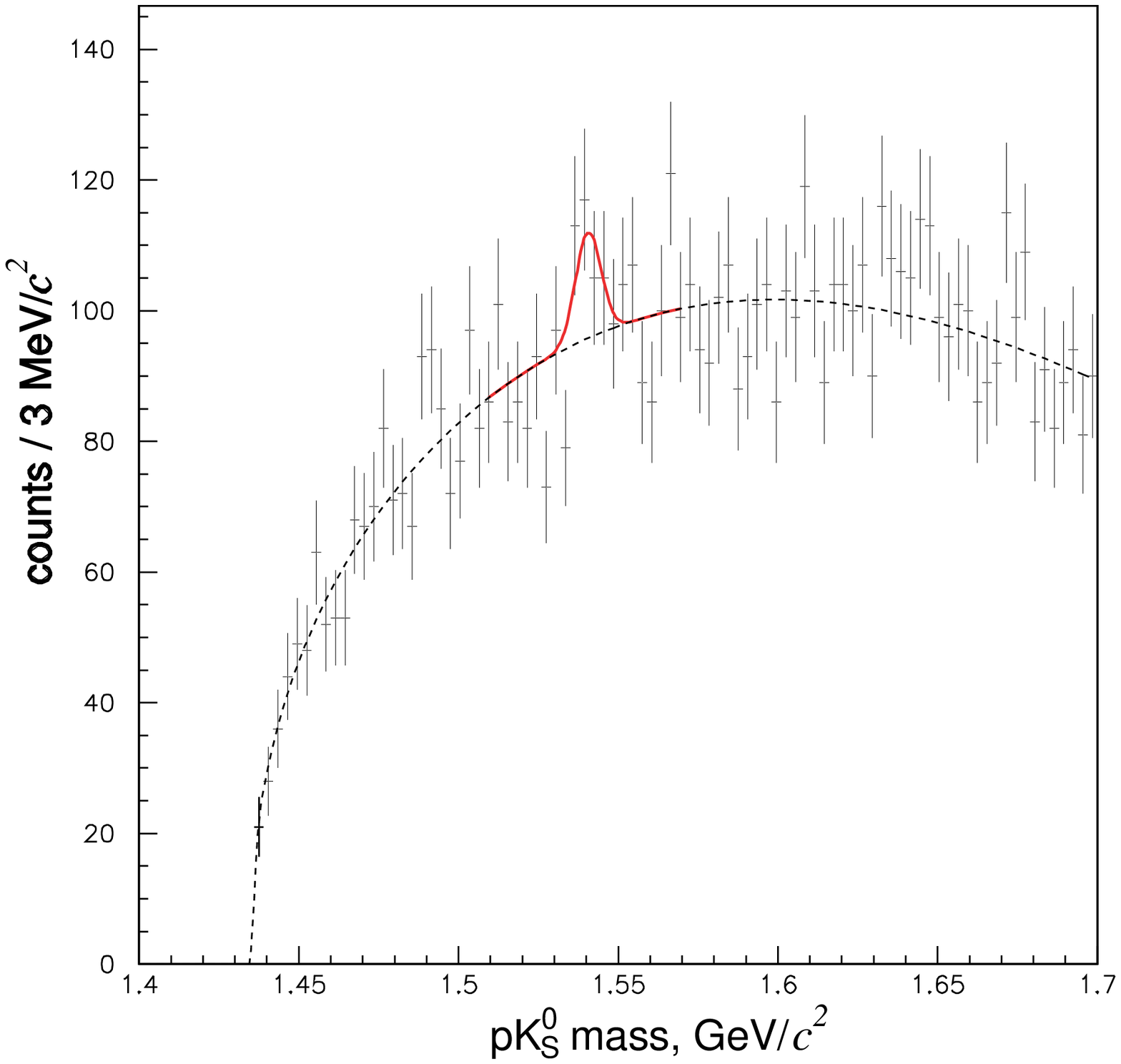,width=0.32\columnwidth}
\epsfig{file=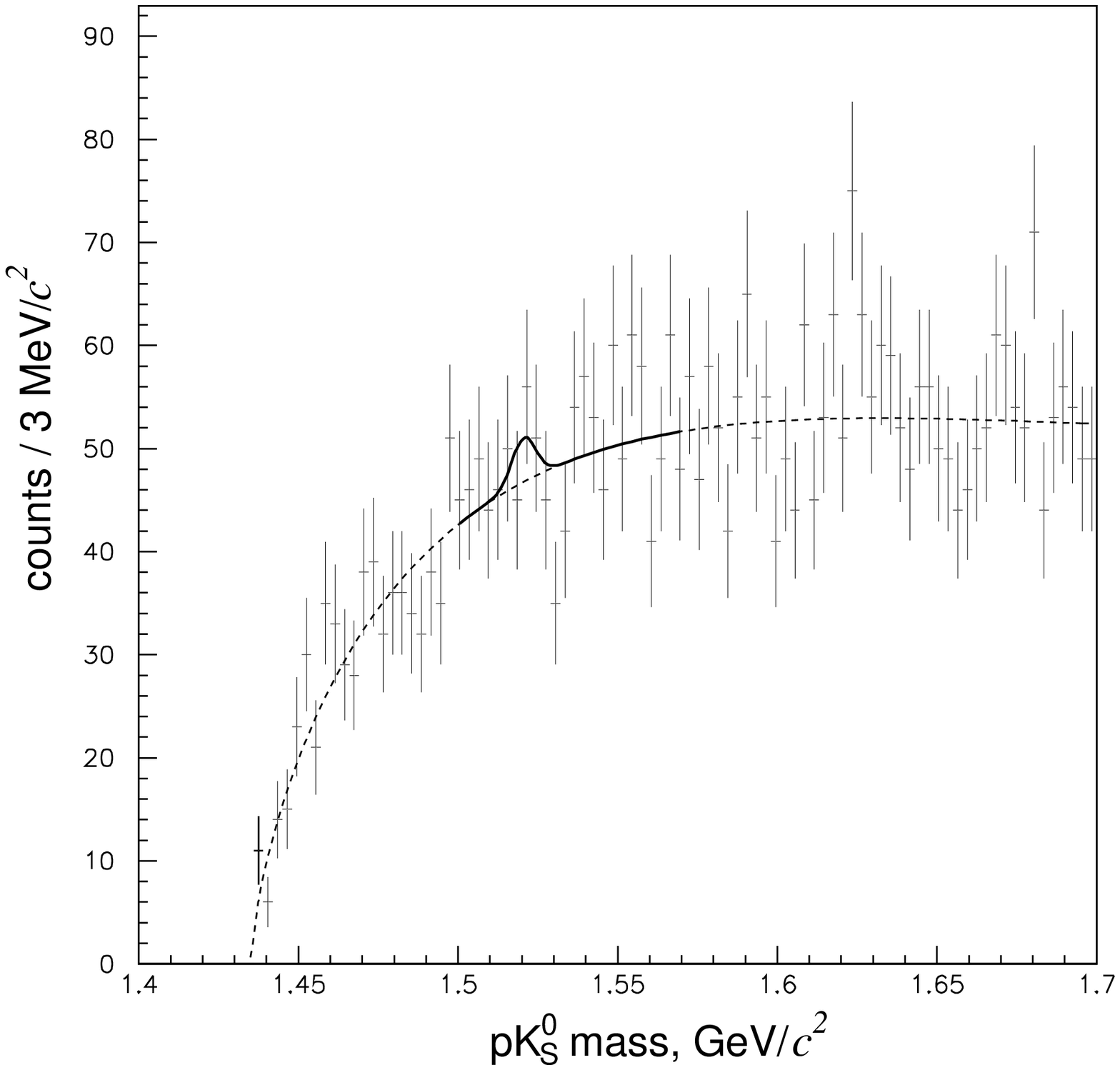,width=0.32\columnwidth}
\epsfig{file=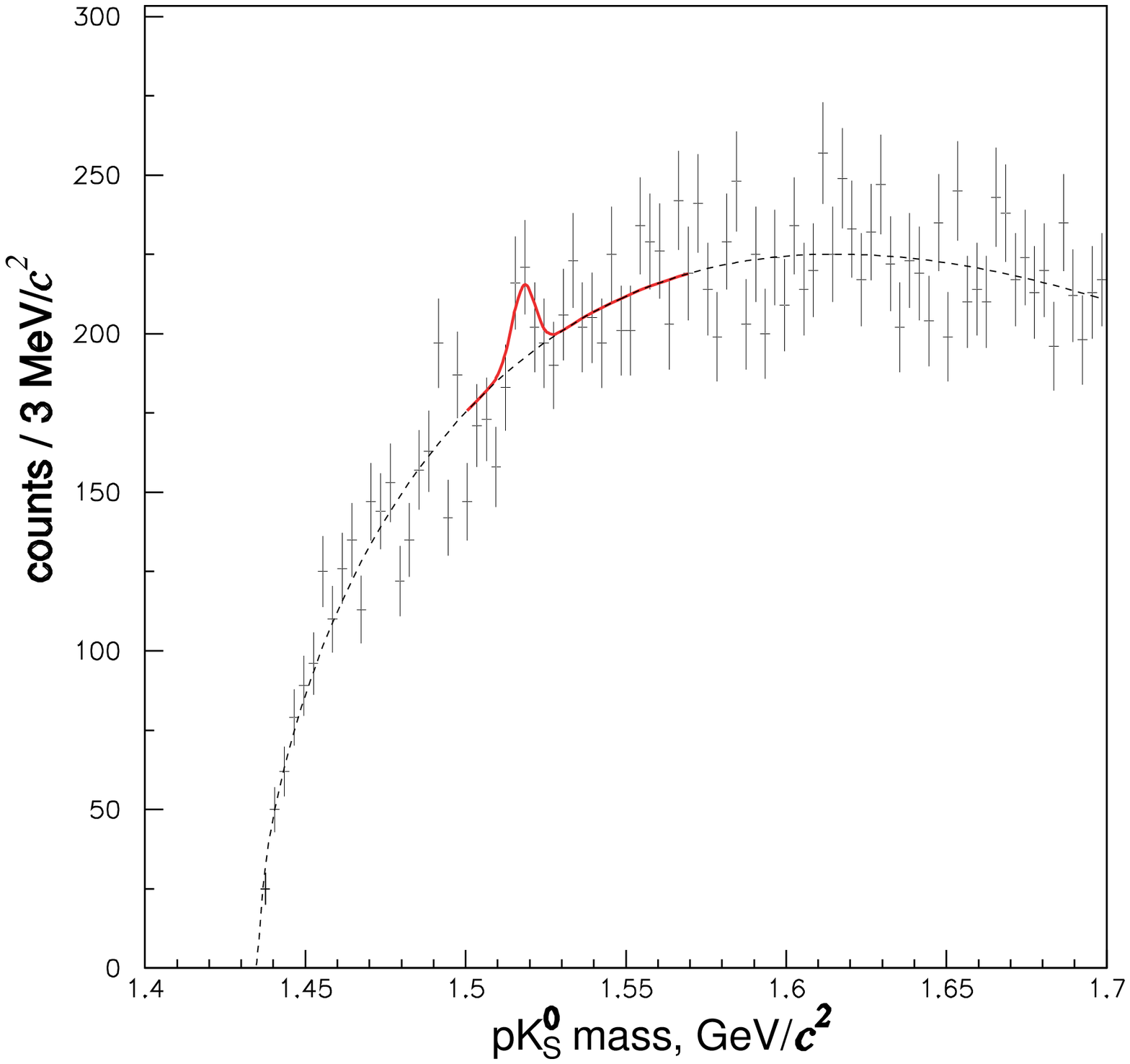,width=0.32\columnwidth}
\put(-330,95){a)}
\put(-210,95){b)}
\put(-90,95){c)}
\caption{The \pKns\  invariant mass distributions for the a) carbon,
b) titanium and c) tungsten data sets. The full line indicates a fit with
a Gaussian and fixed background in the region with the highest upper limit. 
\label{fig:maxfluc} }
\end{figure}

Searching the carbon sample within the mass interval indicated in
Fig.~\ref{fig:prK0s}b,
the largest positive fluctuation of 54$\pm$23 events can be
found at a mass of 1541~\MeV\  (see Fig.~\ref{fig:maxfluc}a).
For the Ti sample (Fig.~\ref{fig:maxfluc}b) we find 13$\pm$22 at a mass of 1521~\MeV\  
and for the W sample (Fig.~\ref{fig:maxfluc}c) 68$\pm$34 at a mass of 1518~\MeV\  ,
respectively. All these peaks are consistent with statistical fluctuations.   
Assuming an atomic mass dependence
of $A^{0.7}$ for the production cross section, the UL(95$\%$) of \Bdsig\ 
varies from 3 to 22~$\mu$b/nucleon for a \The\  mass between 1521 and 
1555~\MeV . The upper limits derived from the full data sample are in agreement
with the carbon results (see Table~\ref{tab:Results}).
 
\begin{table}[ht]
\caption{Upper limits (95$\%$) of \Bdsig\ per nucleon for $\Theta^+$
pentaquarks in $\mu$b/N. \label{tab:Results}}
\begin{center}
\begin{tabular}{|l|c|c|c|c|}
\hline 
Mass (\MeV\ )&\multicolumn{2}{|c|}{1530}&\multicolumn{2}{|c|}{1521 - 1555}
\\ \hline
Signal  &  C target & all targets & C target & all targets
\\ \hline
\Thed\        &   3.7     &   4.8       &  3 - 22  &   4 - 16  
\\ \hline
\end{tabular}
\end{center}
\end{table}

In addition, we have evaluated the upper limit on the ratios of  $\Theta^+$
to $\Lambda$ and $\Lambda(1520)$, allowing for a direct comparison
with other experiments.
Assuming $\mathcal{B}$(\Thed )~=~0.25, we obtain the following upper
limits at 95$\%$ CL: $\Theta^+ / \Lambda$ = 0.92$\%$
and $\Theta^+ / \Lambda(1520)$ = 2.7$\%$ at a mass
of 1530~\MeV , using the full data sample. For a $\Theta^+$ mass of
1540~\MeV , the quoted values must be multiplied by $\approx$4. 
Assuming similar production mechanisms, our limits are at variance with the
findings of HERMES~\cite{HERMES} and ZEUS~\cite{ZEUS, Yoshida} experiments
(see Table~\ref{tab:Ratios}).

\begin{table}[ht]
\caption{Relative yields of $\Theta^+$ assuming
$\mathcal{B}$(\Thed )~=~0.25 for a mass value of 1530~\MeV . The HERA-B
results refer to the full data sample.
\label{tab:Ratios}}
\begin{center}
\begin{tabular}{|l|c|c|c|c|}
\hline
Reaction        & $\sqrt{s}$ (GeV) & source &$\Theta^+ / \Lambda$ & $\Theta^+ / \Lambda(1520)$ 
\\ \hline 
pA, y$\approx$0 &  41.6            & HERA-B   &  $<$ 0.0092         & $<$ 0.027
\\ \hline
ed              &  27.6            & HERMES &                   &   1.6 - 3.5
\\ \hline
ep              &   320            & ZEUS &     0.05            & 
\\ \hline
\end{tabular}
\end{center}
\end{table}

\section{Search for \Xinn\ }

\begin{figure}[htb]
\epsfig{file=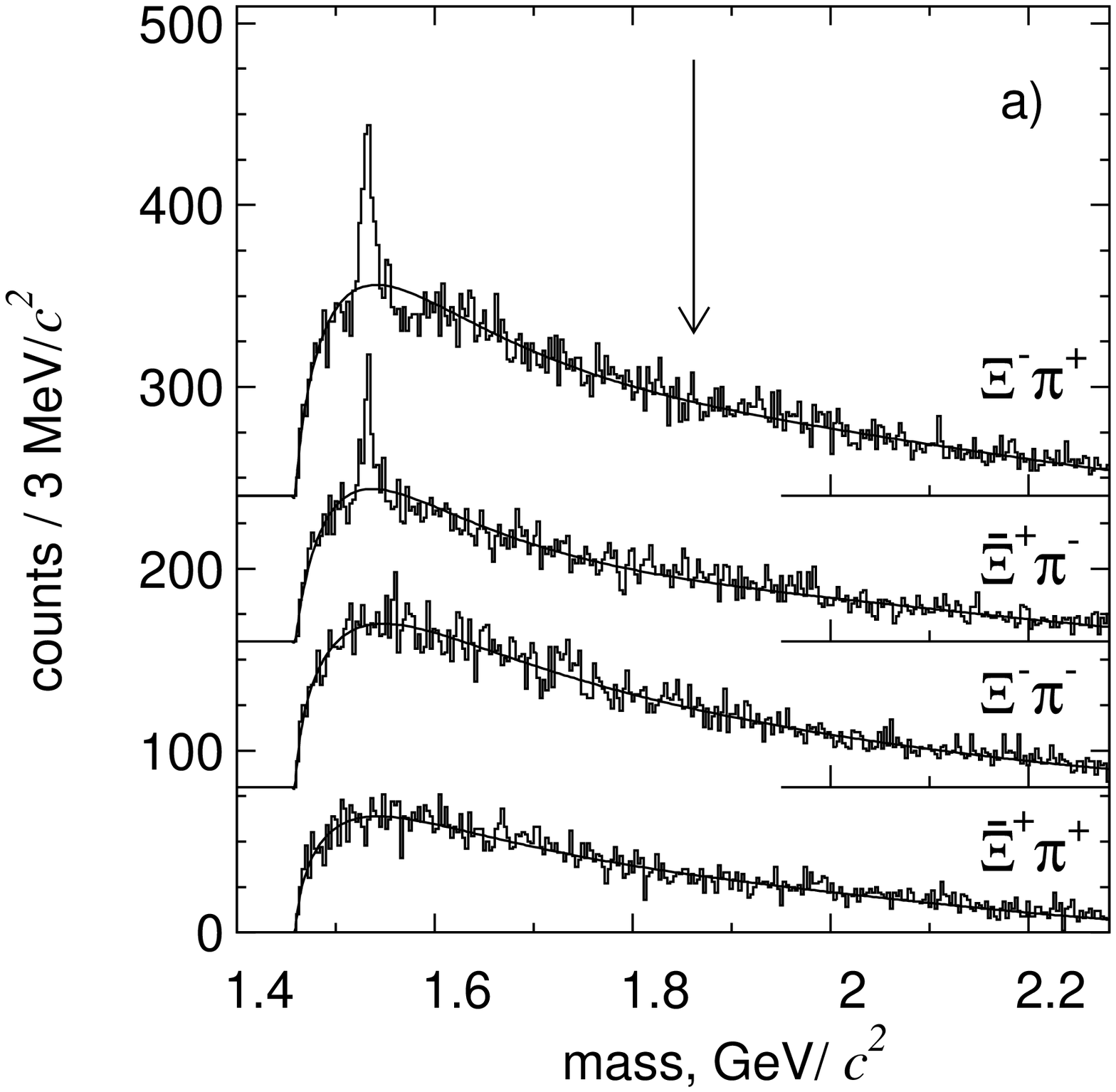,width=0.49\columnwidth}
\epsfig{file=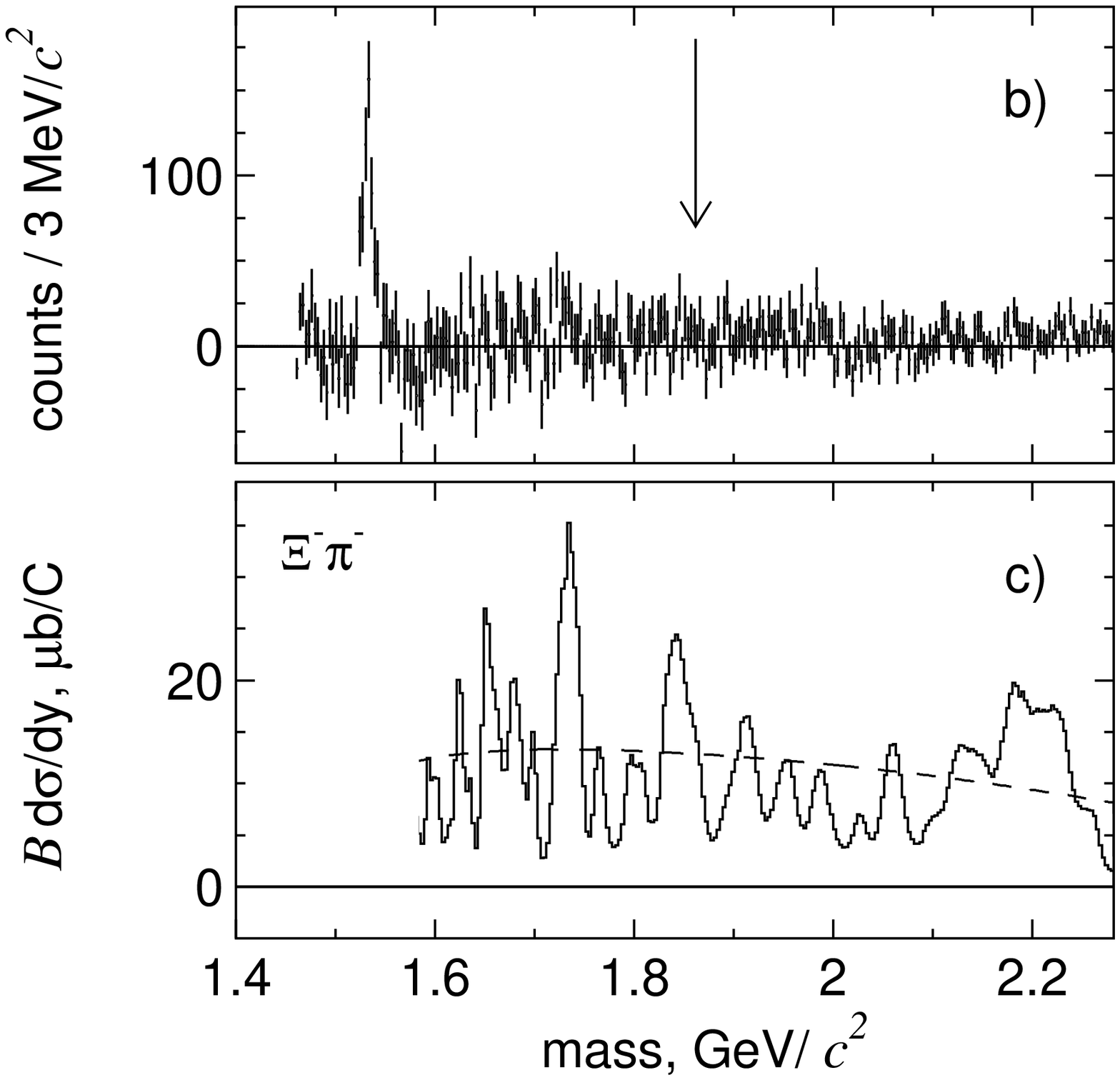,width=0.49\columnwidth}
\caption{The  $\Xi \pi$ invariant mass distributions: a) data from pC collisions
and background estimate (line) in all four charge combinations;
b) sum of all four \Xibb\  spectra with the background subtracted; 
c) upper limit (95$\%$) for the pC production cross section at mid-rapidity;
the dashed line shows our 95$\%$ CL sensitivity. The arrow marks the mass of
1862~\MeV . \label{fig:Xipi}}
\end{figure}

We search for neutral and doubly-charged pentaquark candidates in the
$\Xi \pi$ channels. The pion tracks are requested to come from the primary vertex.
In addition, soft particle identification cuts are applied. The
invariant mass spectra of all four \Xibb\  combinations from our
carbon sample are plotted in Fig.~\ref{fig:Xipi}a. The background estimates obtained
from event mixing normalized to the data are indicated as smooth lines.
The $\Xi^0(1530)$ hyperons show up as a prominent
signal (see Table~\ref{tab:Resonan}) in both neutral charge combinations.
However, none of these mass spectra shows
evidence for the narrow pentaquark candidates at 1862~\MeV\  reported by the
NA49 collaboration~\cite{NA49}. 

The sum of the four spectra is plotted in Fig.~\ref{fig:Xipi}b
and can be compared directly to Fig. 3 of ref.~\cite{NA49}. As described in the
previous chapter, we obtain the upper limit (95$\%$) curve for \Bdsig . 
For the carbon sample, the result for $\Xi^- \pi^-$ combinations is shown
as a solid line in  Fig.~\ref{fig:Xipi}c.
The experimental mass resolution increases from 2.9 to 10.6~\MeV\  in the
mass range considered and is 6.6~\MeV\  at 1862~\MeV . The rapidity interval
selected is $y_{cm}=\pm 0.7$. The dashed line indicates the experimental
sensitivity. The upper limits are summarized in Table~\ref{tab:Result2}.

\begin{table}[ht]
\caption{Upper limits (95$\%$) of \Bdsig\ per nucleon for $\Xi(1862)$
pentaquarks in $\mu$b/N. \label{tab:Result2}}
\begin{center}
\begin{tabular}{|l|c|c|}
\hline
Signal  &  C target & all targets
\\ \hline 
$\Xi^{--}(1862) \rightarrow \Xi^- \pi^-$ &   2.5   &    2.7
\\       
$\Xi^{0}(1862) \rightarrow \Xi^- \pi^+$ &   2.3    &    3.2
\\       
$\overline{\Xi}^{++}(1862) \rightarrow \overline{\Xi}^+ \pi^+$ &   0.85  &  0.94
\\
$\overline{\Xi}^{0}(1862) \rightarrow \overline{\Xi}^+ \pi^-$  &   3.1  &  3.1
\\ \hline
\end{tabular}
\end{center}
\end{table}

The upper limits (95$\%$) on the cross section ratios $\Xi^{--} / \Xi^{0}(1530)$ and
$\Xi^{--} / \Xi^-$ are collected in Table~\ref{tab:Ratios2} together with
limits measured by ZEUS~\cite{ZEUS3} and CDF~\cite{CDF}.
In order to compare our limit with the findings
of NA49, we take the estimated number of \Ximm\  candidates from ref.~\cite{NA49}
and the estimated number (150) of $\Xi^{0}(1530)$ events from the same
data set~\cite{Fisch}.
Assuming the relative efficiencies for \Ximm\ and $\Xi^{0}(1530)$ are similar for
both HERA-B and NA49 detectors, we obtain a cross section ratio of  
$\Xi^{--} / \Xi^{0}(1530) \approx $ 0.18/$\mathcal{B}$ for the NA49 signal which
is in contradiction to our upper limit.

\begin{table}[ht]
\caption{Relative yields of \Xin . The HERA-B results refer to the full data sample.
\label{tab:Ratios2}}
\begin{center}
\begin{tabular}{|l|c|c|c|c|}
\hline
Reaction        & $\sqrt{s}$ (GeV) & source &$\Xi^{--} / \Xi^-$ & $\Xi^{--} / \Xi^{0}(1530)$ 
\\ \hline 
pA, y$\approx$0 &  41.6            & HERA-B   &  $<$ 0.03/$\mathcal{B}$& $<$ 0.04/$\mathcal{B}$
\\ \hline 
pp              &  17.0            & NA49   &                      & 0.18/$\mathcal{B}$
\\ \hline
ep              &   320            & ZEUS   &    &  $<$ 0.28/$\mathcal{B}$
\\ \hline
pp              &   1960           & CDF    &    &  $<$ 0.04/$\mathcal{B}$
\\ \hline
\end{tabular}
\end{center}
\end{table}

\section{Summary}

In conclusion, HERA-B has found no evidence for narrow pentaquarks
decaying into \pKns\ or \Xibb\  final states.
We have set upper limits at 95$\%$ CL on the production cross sections
\Bdsig\ at mid-rapidity assuming the width of these states to be less than
our experimental resolution of $\approx$5~\MeV  . Searching the full data sample
for \Thed\  decays, the limit varies from 4 to 16~$\mu$b/nucleon for a mass
between 1521 and 1555~\MeV . Assuming comparable production mechanisms, our
limits of $\Theta^+ / \Lambda < $ 0.92$\%$ and
$\Theta^+ / \Lambda(1520) <$ 2.7$\%$ contradict the results of ZEUS and
HERMES experiments, respectively.
Analyzing the $\Xi \pi$ channels, we determine upper limits \Bdsig\  for \Xin\    
pentaquark candidates. Combining all three data samples, we obtain for a
narrow resonance at a mass of 1862~\MeV ,  2.7, 3.2, 0.94 and 3.1~$\mu$b/nucleon
for $\Xi^- \pi^-$, $\Xi^- \pi^+$, $\overline{\Xi}^+ \pi^+$ and
$\overline{\Xi}^+ \pi^-$ final states, respectively. The upper limit (95$\%$) on
the cross section ratio $\Xi^{--} / \Xi^{0}(1530)$ of 0.04/$\mathcal{B}$
is inconsistent with the evidence published by NA49 for such a resonance.

\end{document}